\documentclass[prd,preprintnumbers,nofootinbib]{revtex4} 
\usepackage{graphicx} 
\usepackage{amsmath}
\usepackage{epstopdf}
\usepackage{amsfonts,amsbsy}
\usepackage{amssymb}
\newcommand{\as}{\alpha_S}

\def\gsim{ \,\, \vcenter{\hbox{$\buildrel{\displaystyle >}\over\sim$}}
 \,\,}

\def\be{\begin{equation}}
\def\ee{\end{equation}}
\def\bea{\begin{eqnarray}}
\def\eea{\end{eqnarray}}

\usepackage{color}

\begin{document}

\title{\bf Gluon saturation in $pA$ collisions at the LHC:\\ predictions for hadron multiplicities}

\author{Adrian Dumitru$^{a,b}$, Dmitri E. Kharzeev$^{c,d}$, Eugene M. Levin$^{e,f}$ and Yasushi Nara$^g$}
\affiliation{
$^a$ RIKEN BNL Research Center, Brookhaven National
  Laboratory, Upton, NY 11973, USA\\
$^b$ Department of Natural Sciences, Baruch College, CUNY,
17 Lexington Avenue, New York, NY 10010, USA\\
$^c$ Department of Physics and Astronomy, Stony Brook University, Stony Brook, NY 11794, USA\\
$^d$ Department of Physics, Brookhaven National
  Laboratory, Upton, NY 11973, USA\\
$^e$ \, Department of Particle Physics, School of Physics and Astronomy,
Tel Aviv University, Tel Aviv, 69978, Israel\\
$^f$\, Departamento de F\'\i sica,
Centro de Estudios Subat$\acute{o}$micos,
Universidad T$\acute{e}$cnica Federico Santa Mar\'\i a,\\ and
Centro Cienti´ıfico-Tecnol$\acute{o}$gico de Valpara\'\i so,
Casilla 110-V,  Valparaiso, Chile\\
$^g$ Akita International University, Yuwa, Akita-city 010-1292, Japan}
\begin{abstract}
The upcoming $p+Pb$ run at the LHC will probe the nuclear gluon
distribution at very small Bjorken $x$ (from $x \sim 10^{-4}$ at
mid-rapidity down to $x \sim 10^{-6}$ in the proton fragmentation
region) and will allow to test approaches based on parton saturation.  Here, we
present the predictions of the KLN model for hadron multiplicities and
multiplicity distributions in $p+Pb$ collisions at a center-of-mass
energy of 4.4~TeV.  We also compare the model to the existing $pp$,
$dA$ and $AA$ data from RHIC and LHC.
\end{abstract}

\maketitle

%---------------------------------------------------------------------------
Very soon, the Large Hadron Collider will record the first data on
$p+Pb$ collisions at the center-of-mass energy of 4.4~TeV. This data
will allow to probe the nuclear gluon distributions at very small
Bjorken $x$: from $x \sim 10^{-4}$ at mid-rapidity down to $x \sim
10^{-6}$ in the proton fragmentation region. Since the QCD evolution
makes parton distributions increase at small $x$, the LHC experiments
will allow to probe the nuclear wave functions at unprecedented parton
densities. These measurements are crucial for testing the current
theoretical approaches to high energy QCD.  \vskip0.3cm Due to the
breaking of scale invariance by quantum effects, QCD possesses a
dimensionful scale $\Lambda_{\rm{QCD}}$ that determines the
characteristic distance $\sim \Lambda_{\rm{QCD}}^{-1}$ at which the
dynamics becomes non-perturbative. The asymptotic
freedom \cite{Gross:1973id,Politzer:1973fx} makes the perturbative
expansion valid only if a hard external scale $Q^2 \gg
\Lambda_{\rm{QCD}}^2$ is present.  Multiparticle production in hadron
collisions is dominated by soft interactions and so in general is not
amenable to the weak coupling treatment.  However when the density of
partons in the transverse plane $Q_s^2$ becomes large compared to
$\Lambda_{\rm{QCD}}^2$, it regularizes the infrared behavior of the parton
transverse momentum distributions at the ``saturation momentum" \cite{GLR} $Q_s$ and
thus prevents the running coupling of QCD from growing large,
$\alpha_s(Q_s) = g^2/4\pi \ll 1$ \cite{GLR,hdQCD,MV}.  The gluon field
$A$ in this weak coupling regime has a large occupation number, $A
\sim 1/g > 1$ and can be treated as a classical "Color Glass
Condensate" (CGC) \cite{MV,BK,JIMWLK}.  \vskip0.3cm

While the complete theory of multi-particle production based on the
ideas outlined above is still being developed, its main ingredients
are clear and can serve as the basis for phenomenology. This was the
motivation for the KLN model
\cite{Kharzeev:2000ph,Kharzeev:2001gp,Kharzeev:2001yq} combining the
Glauber approach to proton-nucleus and nucleus-nucleus collisions (for
a complete set of formulae see e.g. \cite{Kharzeev:1996yx}) with a
simple ansatz for the unintegrated parton distributions that accounts
for the existence of a new dimensionful scale -- the saturation
momentum.  The KLN model was successful in describing the RHIC data
\cite{Brahms,Phenix,Phobos,Star} on the centrality and rapidity
dependence of charged hadron production in heavy ion collisions. The
predictions for Pb Pb and p Pb collisions at the LHC were made in
\cite{Kharzeev:2004if}. The comparison to the first LHC data
\cite{Aamodt:2010cz} on hadron production in Pb-Pb collisions revealed
that while the KLN model describes the centrality dependence rather
well, the overall normalization exceeds the observed one by about
10-15 $\%$. This implies that the energy dependence of the saturation
momentum assumed in \cite{Kharzeev:2004if} was slightly too
steep\footnote{While it is evident that the model has to be refined,
  let us put this discrepancy in perspective by noting that some of
  the early pre-RHIC predictions for the LHC that did not take into
  account the concepts of parton saturation and coherence
  overestimated the measured hadron multiplicity by almost an order of
  magnitude.}.  \vskip0.3cm

Regarding pA collisions, we also have to remember that the
number of ``participants" (the nucleons that underwent at least one
inelastic interaction) in this case is much smaller than in A A
collisions, and that fluctuations are much more important. Therefore
a Monte-Carlo (MC) based formulation~\cite{MCKLN} of the numerical
integration of the KLN
model~\cite{Drescher:2006pi} can be expected to provide more accurate
predictions. Indeed, the MC method leads to a better agreement between
the data and the model prediction \cite{Kharzeev:2002ei} in d Au
collisions at RHIC. The MC based KLN model~\cite{MCKLN} has been
used to generate initial conditions for the hydrodynamical description
of collective flow, see
e.g.~\cite{Hirano:2005xf,Hirano:2009ah,Song:2010mg}.  \vskip0.3cm

The goal of this letter is to provide updated predictions for p Pb
collisions at the LHC. Let us explicitly list the differences between
the present and the previous \cite{Kharzeev:2004if} papers: i) we
consider the c.m.s.\ energy of the forthcoming p Pb run -- 4.4~TeV; ii)
we have reduced the intercept describing the energy dependence of
saturation momentum by $\sim 20\%$; iii) we employ the MC method of
evaluating the number of participants. Of course, after making these
changes we have to make sure that the RHIC data is still adequately
described -- therefore we present the comparison to the RHIC AA and dA
as well. While these changes may seem insignificant, the p Pb LHC data
present a chance to test saturation ideas,
and this requires quantitative predictions made to the best of our
current knowledge.  \vskip0.3cm Let us briefly recall the basic
ingredients of the KLN approach; for details, see
\cite{Kharzeev:2001gp,Kharzeev:2004if}. The multiplicity per unit
rapidity
\be  \label{eq:dNdy}
\frac{dN}{dy} = {K \over S}\ \ \int d p_t^2 \left( E {d \sigma \over d^3 p} \right) 
=  {K \over S}\ \ {4 \pi N_c \as \over N_c^2 - 1}\
\,\int^{\infty}_0\,\,\frac{
d\,p^2_t}{p^4_t}\,\,x_2G_{A_2}(x_2,p^2_t)\,\, x_1G_{A_1}(x_1,p^2_t)\,\,,
\ee
is evaluated using the gluon density obtained from a simple ansatz for the unintegrated gluon distribution \cite{Kharzeev:2001gp} encoding the saturation phenomenon:
\be
xG(x,p_t^2) = 
\left\{
\begin{array}{l c r}
 \frac{S}{\alpha_s(Q_s)} \; p_t^2 \;\left(1- x \right)^4 & , & p_t<Q_s(x)\\
 \frac{S}{\alpha_s(Q_s)} \; Q_s^2 \;\left(1-x \right)^4 & , & p_t>Q_s(x)
\end{array}\right.
\label{eq:xG}
\ee
where $x = x_1$ or $x_2$, with $x_{1,2} = (p_t / W) e^{\pm y}$; the
+(-) sign in the exponent applies to the projectile (target), and
$W\equiv \sqrt{s}$ is the c.m.s. energy. The factor $S$ in
eq.~(\ref{eq:dNdy}) is the transverse area involved in the collision
(see below). The normalization factor $K$ describes the conversion of
partons to hadrons and is determined by a global fit to $pp$ data at
various energies, and to $d+Au$ data from RHIC.  \vskip0.3cm To
describe the running of QCD coupling, we use the $\beta$-function in
the one-loop approximation with $N_f=3$ light quark flavors and
$\Lambda_{\rm QCD}^2=0.05$~GeV$^2$ but assume that the coupling
freezes at $\alpha_{\rm max}=0.52$ \cite{Dokshitzer:2004tp}:
\be
\alpha_s(Q^2) = \mathrm{min}\left[
\frac{12\pi}{27\; \log \frac{Q^2}{\Lambda_{\rm QCD}^2}} \, ,\,
\alpha_{\rm max} \right]~~~~,~~~(Q^2\ge\Lambda_{\rm QCD}^2).
\ee
The factor of $\alpha_s(Q^2)$ in the integral (\ref{eq:dNdy}) is
evaluated at the scale $p_t^2$, if this is the largest scale, or else
at the lower of $Q_{s,P}^2(y)$ and $Q_{s,T}^2(y)$. The saturation
momenta are defined as
\be \label{eq:Qs2}
Q_s^2(y) = Q_0^2 \; N_{\rm part}\; \left(x_0 \frac{W}{Q_0} e^{\mp y} \right)^{\bar\lambda} ~,
\ee
where again the +(-) sign in the exponent applies to the projectile
(target). We fix the parameters to $Q_0=0.6$~GeV, $x_0=0.01$, and
$\bar\lambda=0.205$. In the midrapidity region of collisions at RHIC
energy, this results in a gluon saturation momentum
$Q_s\simeq0.68$~GeV for a proton.
On account of the large radius of the deuteron, we have used
$N_{\rm part,P}=1$ in~(\ref{eq:Qs2}) in this case assuming that the
parton substructure of the nucleon in the deuteron is not
modified. For minimum bias $d+Au$ collisions we multiply $dN/dy$
by a factor of 1.52 which is our estimate for the corresponding
equivalent number of $p+Au$ collisions at an energy of $W=200$~GeV.
For $pp$ collisions we choose the effective area $S_{pp}\simeq 0.7\;
S_{pA}$ somewhat smaller than for $pA$ collisions, as suggested by the
data. This may be an indication that in proton-proton collisions only
part of the proton takes part in the interaction. On the other hand,
the large nucleus makes all of the proton's constituents to interact.
\vskip0.3cm 
To evaluate the pseudo-rapidity distributions,
Eq.~(\ref{eq:dNdy}) needs to be rewritten using the transformation
\be \label{eq:y_eta}
y(\eta) = \frac{1}{2} \log\frac
{\sqrt{\cosh^2\eta + \mu^2} + \sinh\eta}
{\sqrt{\cosh^2\eta + \mu^2}-\sinh\eta}
\ee
with the Jacobian
\be \label{eq:J_eta}
J(\eta) = \frac{\partial y}{\partial\eta} = 
\frac{\cosh\eta}{\sqrt{\cosh^2\eta + \mu^2}}~.
\ee
The scale $\mu^2(W)$ is allowed to exhibit a weak energy dependence
according to
\be  \label{eq:mu_eta}
\mu(W) = \frac{0.24}{0.13 + 0.32\;W^{0.115}}~,
\ee
with $W$ expressed in units of TeV. This parameterization reproduces
approximately the ``shoulder'' structure of $dN/d\eta$ observed in
symmetric $pp$ collisions. We did not modify $\mu(W)$ for the case of
$pA$ collisions.
\vskip0.3cm

The multiplicity discussed above represents the {\it average}
multiplicity $\langle N_{\rm ch}\rangle$ observed in collisions with a
fixed number of participants. In experiment, the multiplicity fluctuates both due to the fluctuations in the number of participants and due to ``intrinsic" fluctuation at fixed number of participants. To model the ``intrinsic'' fluctuations
of the number of produced particles we consider the multiplicity
(per unit rapidity) as a random variable distributed according to a
negative binomial distribution,
\be
P(N_{\rm ch}) = \frac{\Gamma(k+n)}{\Gamma(k)\Gamma(n+1)}
\frac{\langle N_{\rm ch}\rangle^{N_{\rm ch}} k^k}
{\left(\langle N_{\rm ch}\rangle+k \right)^{N_{\rm ch}+k}} ~.
\ee
The quantity $k$ which characterizes the fluctuations in the
saturation approach has been estimated as
be~\cite{Gelis:2009wh,Tribedy:2010ab} 
\be
k = \kappa \, \frac{N_c^2-1}{2\pi}\, Q_s^2(y,W) \; \sigma_k(W)~.
\ee
In our numerical estimates we have assumed that $\sigma_k(W) = \sigma_{\rm
  in}(W)/10$ is proportional to the inelastic $pp$ cross
section, and that $Q_s$ is the saturation scale of the proton. We find that the value of $\kappa$ which
describes best the multiplicity distributions in $pp$ collisions is about $\kappa\simeq 0.05$.

\vskip0.3cm
All observables for $pA$ collisions finally need to be averaged also
over an ensemble of $N_{\rm part,A}$, which enters through
eq.~(\ref{eq:Qs2}). We obtain the number of participants in the
heavy ion target from a Monte-Carlo Glauber simulation\footnote{On the
  other hand, in AA collisions the fluctuations of $N_{\rm part}$ do
  not affect the multiplicity strongly; we have calculated $N_{\rm part}$
  directly, in a ``mean field approximation'', from a nuclear
  Woods-Saxon distribution.}: assume a
uniformly distributed random number $0<\nu<1$ and let
\be
N_{\rm part,A}(\vec{b}) = \sum\limits_{i=1\cdots A}
\Theta\left(P(\vec{b}-\vec{r}_i) -\nu_i\right)~.
\ee
Here, $b$ is the impact parameter of the $p+A$ collision, i.e.\ the
transverse distance of the proton from the center of the target
nucleus; it is a random variable with the probability density $b\;
db$. The set $\{ \vec{r}_i \}$ corresponds to the coordinates of the
nucleons in the target which are picked randomly according to a
Woods-Saxon distribution. Finally, $P(r)$ denotes the interaction
probability of two nucleons separated by a transverse distance $r$;
for simplicity, here we assume ``hard sphere'' nucleons:
\be
P(r)=\Theta\left(\sqrt\frac{\sigma_{\rm in}(W)}{\pi}-r\right)~.
\ee
We use the measured values $\sigma_{\rm in}(s)=42$, 52, 60, 65.75,
70.45~mb at $W=200$, 900, 2360, 4400, 7000~GeV, respectively. 

\vskip0.3cm

\begin{figure}[htb]
\begin{center}
\includegraphics[width=0.45\textwidth]{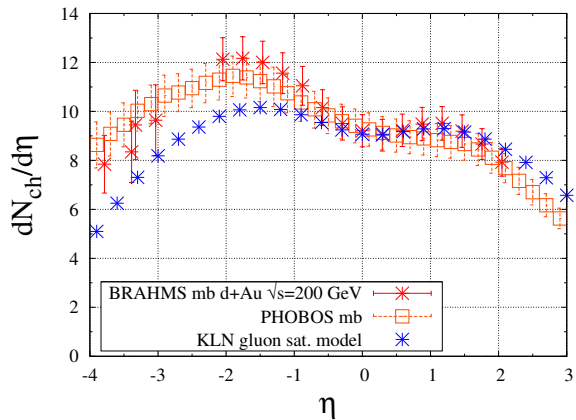}
\end{center}
\caption{
Rapidity distribution of charged particles in minimum bias $d+Au$ collisions at
$W=200$~GeV.
PHOBOS and BRAHMS data from refs.~\cite{Back:2004mr,Arsene:2004cn}.
} \label{fig:dAu200}
\end{figure}

Let us now present and describe our results. First we re-check the
model against the RHIC data. Fig.~\ref{fig:dAu200} shows the
comparison to the d Au data; in the range $-1 < \eta < 2$ the
agreement is satisfactory.  Note that at $\eta \gsim 2$ the saturation
momentum of the projectile becomes small and so the validity of the
saturation approach is questionable at best. Also, in the
fragmentation region of the nucleus one would have to account for the
contribution from valence quarks to improve agreement with the data.

\begin{figure}[htb]
\begin{center}
\includegraphics[width=0.45\textwidth]{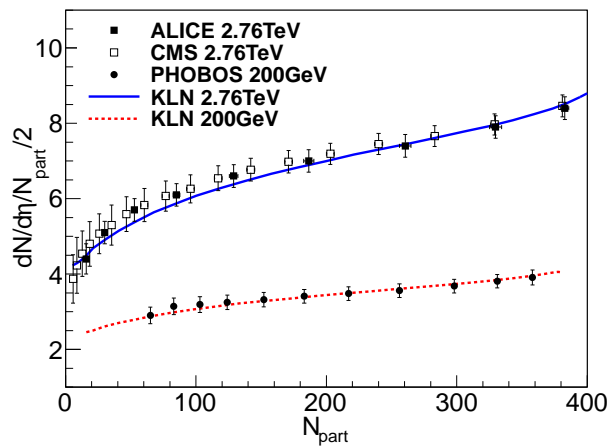}
\end{center}
\caption{Centrality dependence of the charged hadron multiplicity at
  $\eta=0$ in $AuAu$ collisions at $W=200$ GeV \cite{Phobos} and
  $PbPb$ collisions at $W=2.76$ TeV \cite{Aamodt:2010cz,
    Chatrchyan:2011pb}}.
\label{fig:AA}
\end{figure}

\begin{figure}[htb]
\begin{center}
\includegraphics[width=0.45\textwidth]{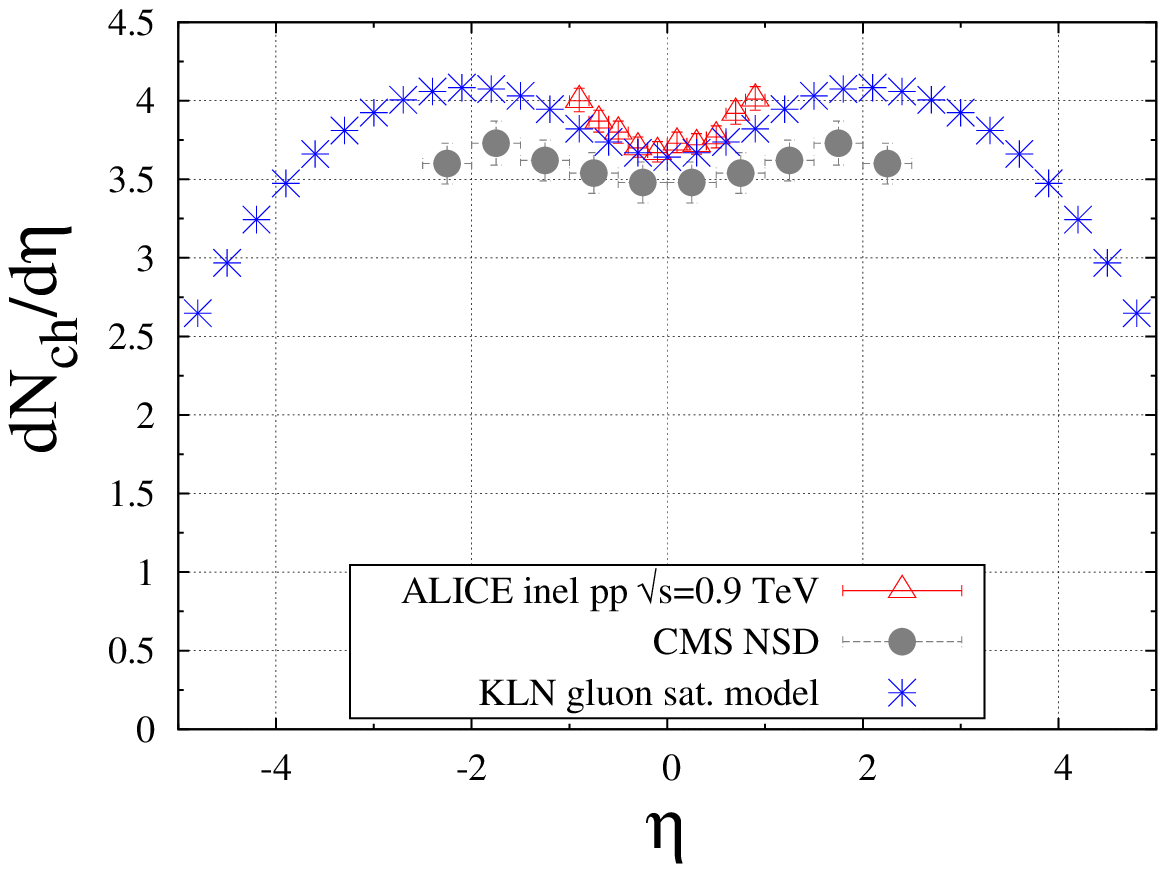}
\includegraphics[width=0.47\textwidth]{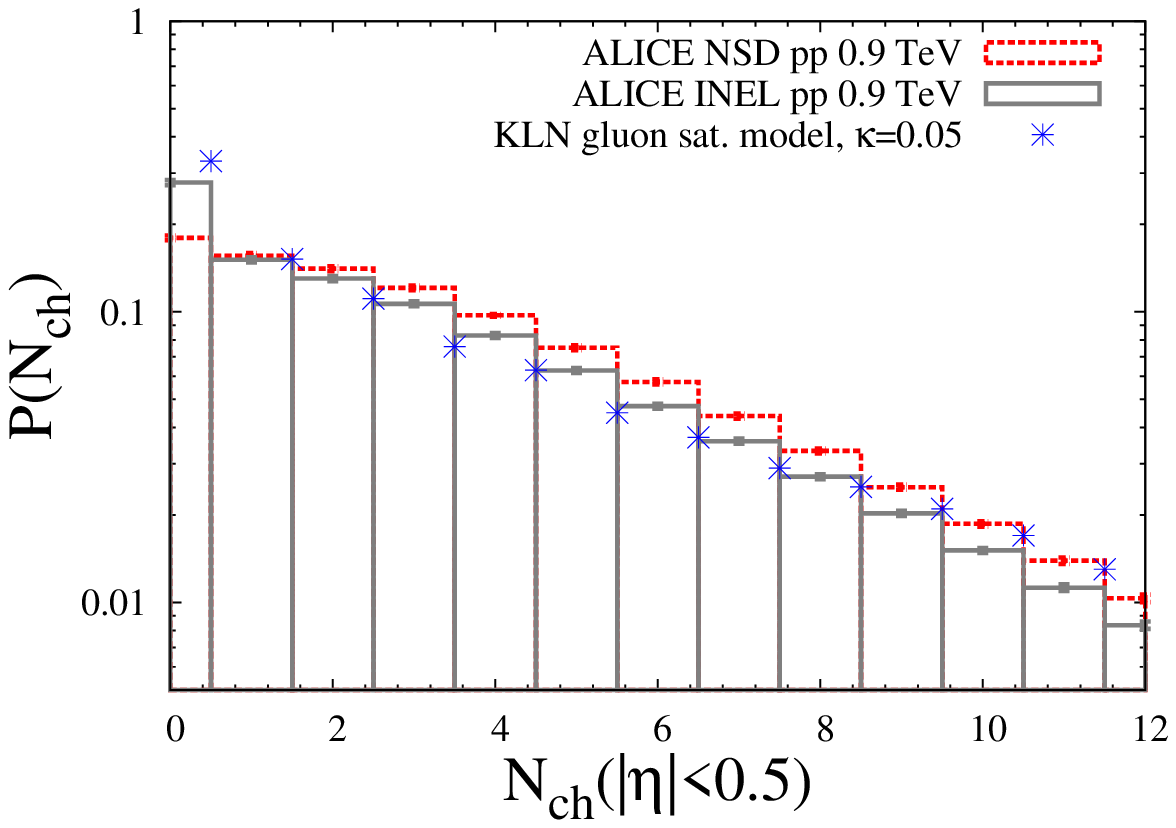}
\end{center}
\caption{
Left: rapidity distribution of charged particles in $pp$ collisions at
$W=900$~GeV. Right: Charged particle multiplicity distribution.
ALICE and CMS data from refs.~\cite{ALICEpp,CMSpp}.
} \label{fig:pp900}
\end{figure}

\begin{figure}[htb]
\begin{center}
\includegraphics[width=0.45\textwidth]{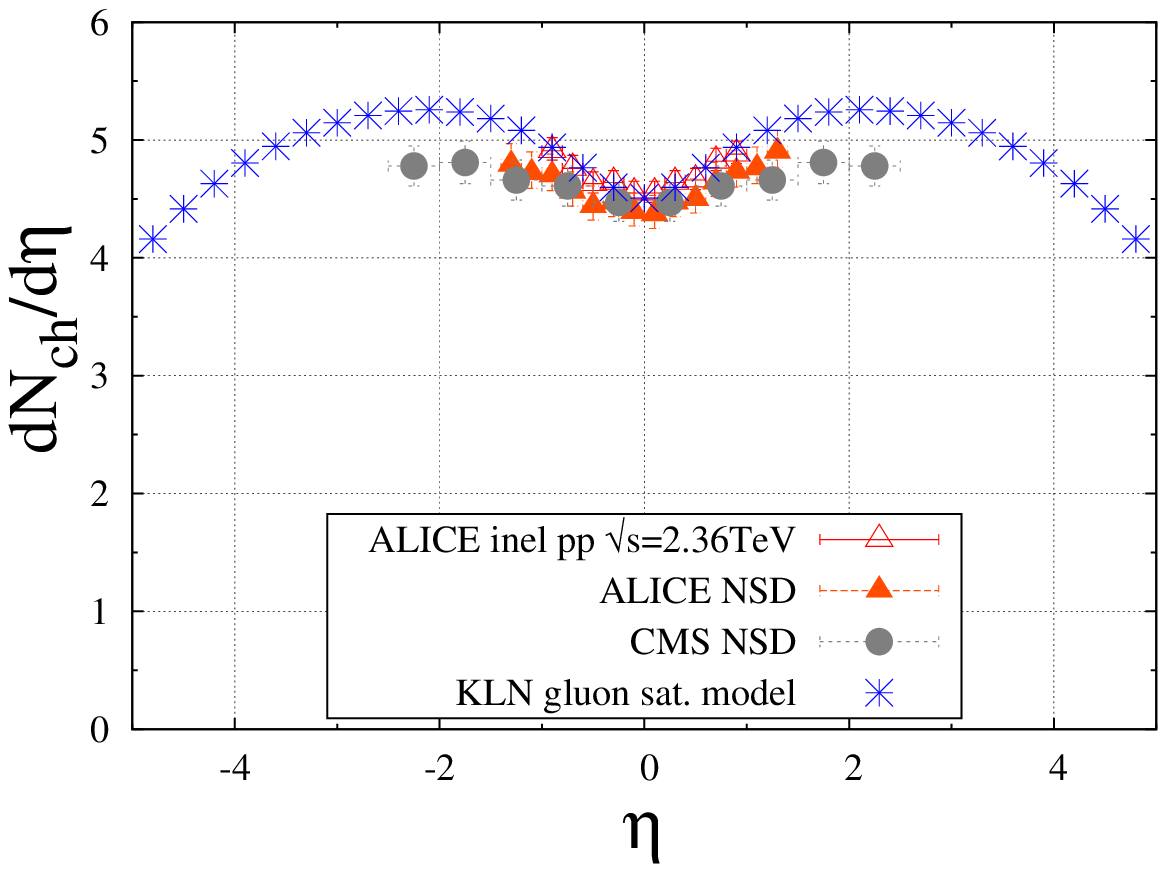}
\includegraphics[width=0.47\textwidth]{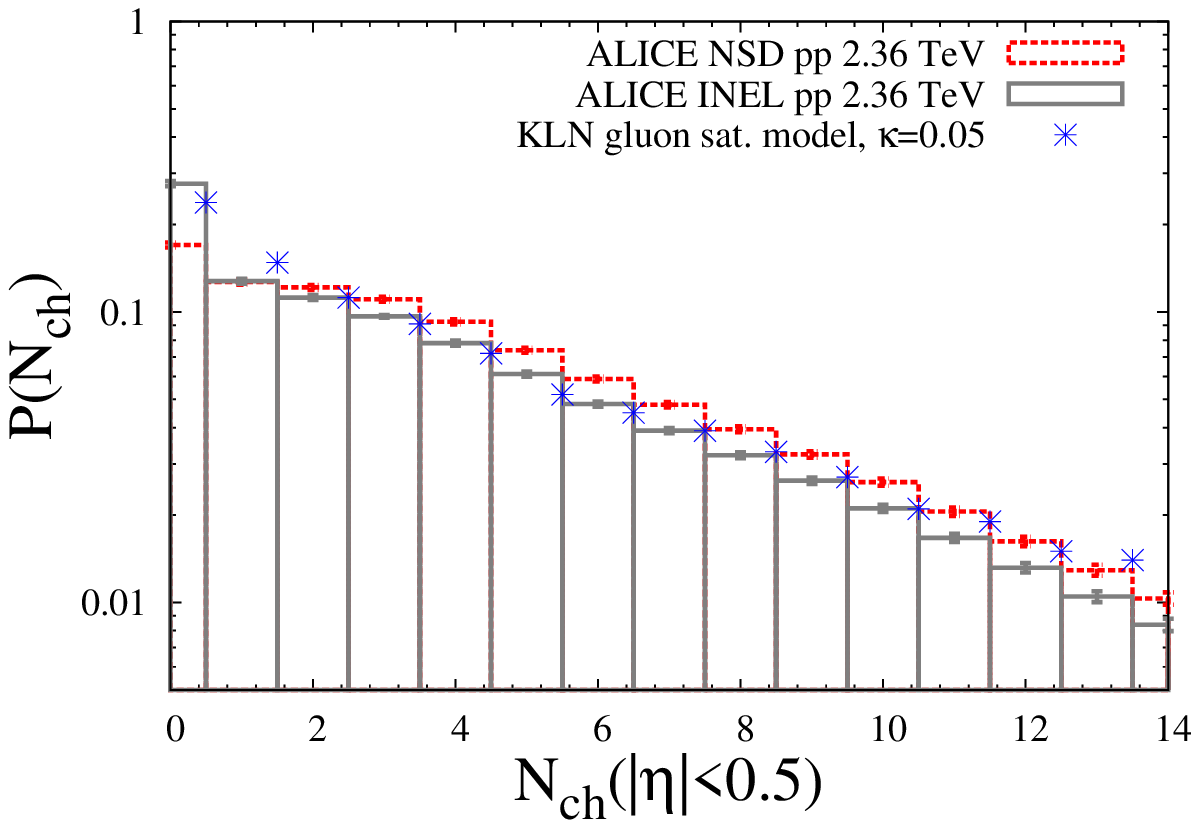}
\end{center}
\caption{
Left: rapidity distribution of charged particles in $pp$ collisions at
$W=2360$~GeV. Right: Charged particle multiplicity distribution.
ALICE and CMS data from refs.~\cite{ALICEpp,CMSpp}.
} \label{fig:pp2360}
\end{figure}

\begin{figure}[htb]
\begin{center}
\includegraphics[width=0.45\textwidth]{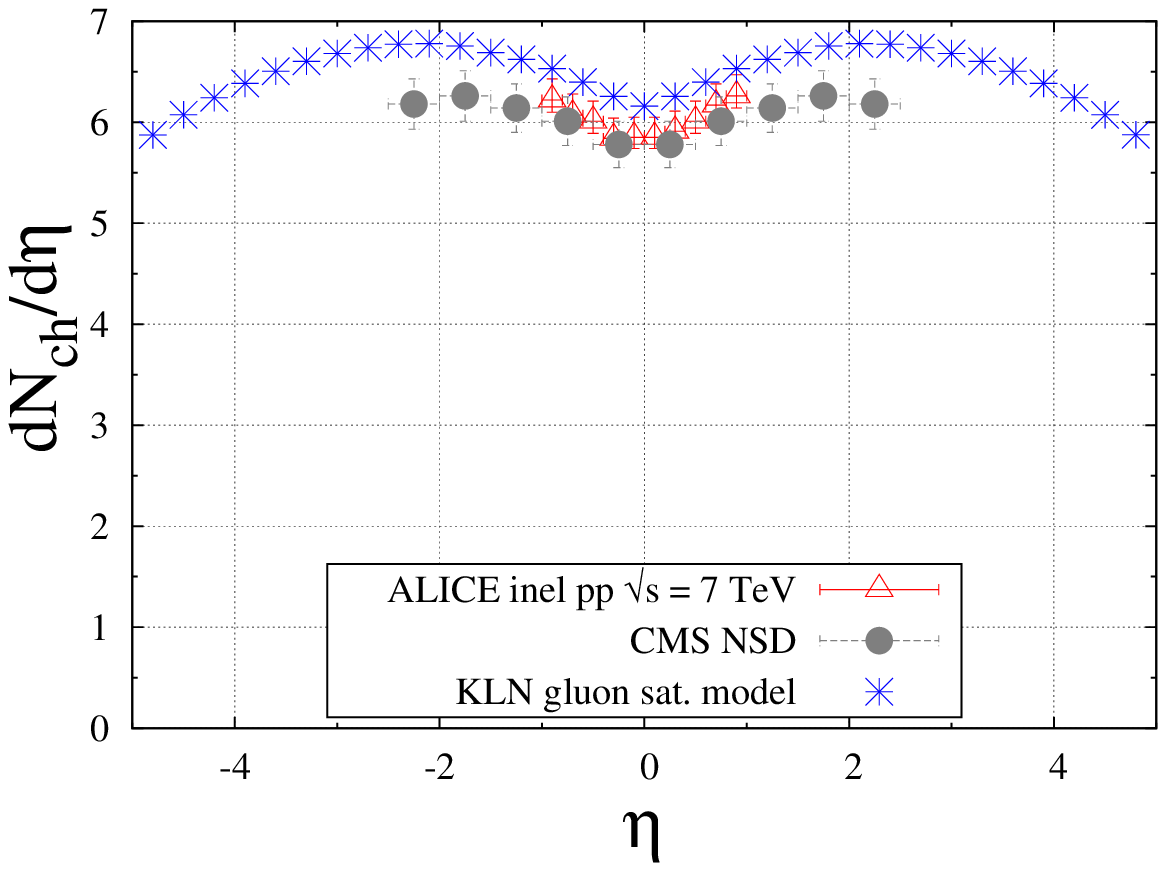}
\includegraphics[width=0.47\textwidth]{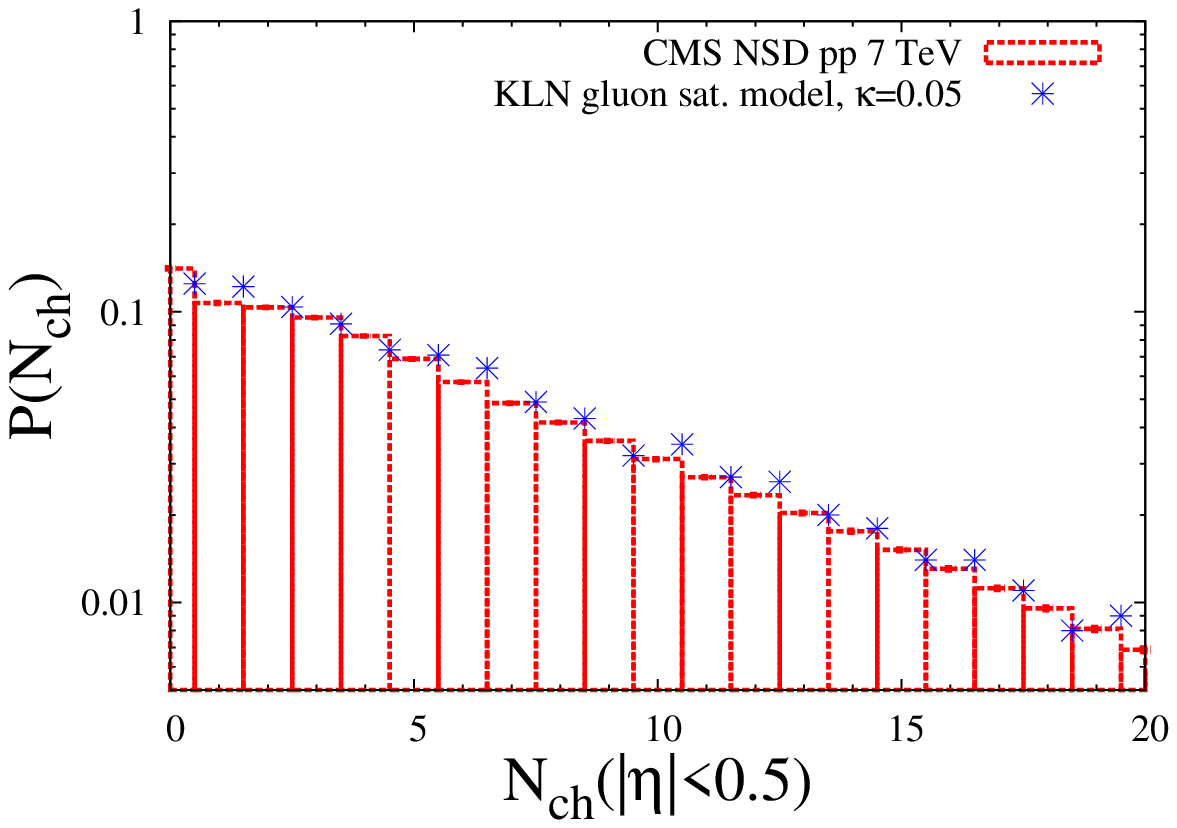}
\end{center}
\caption{
Left: rapidity distribution of charged particles in $pp$ collisions at
$W=7000$~GeV. Right: Charged particle multiplicity distribution.
ALICE and CMS data from refs.~\cite{ALICEpp,CMSpp}.
} \label{fig:pp7000}
\end{figure}

\begin{figure}[htb]
\begin{center}
\includegraphics[width=0.45\textwidth]{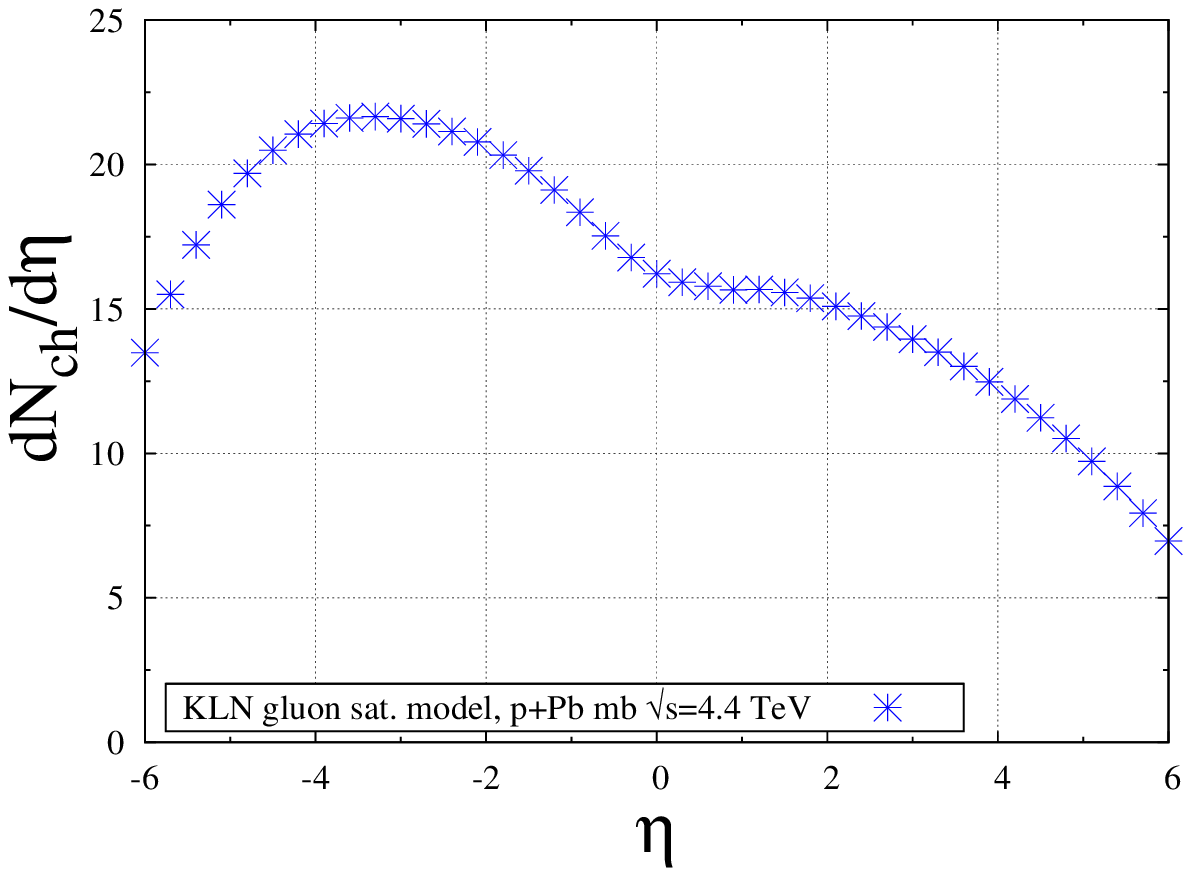}
\includegraphics[width=0.47\textwidth]{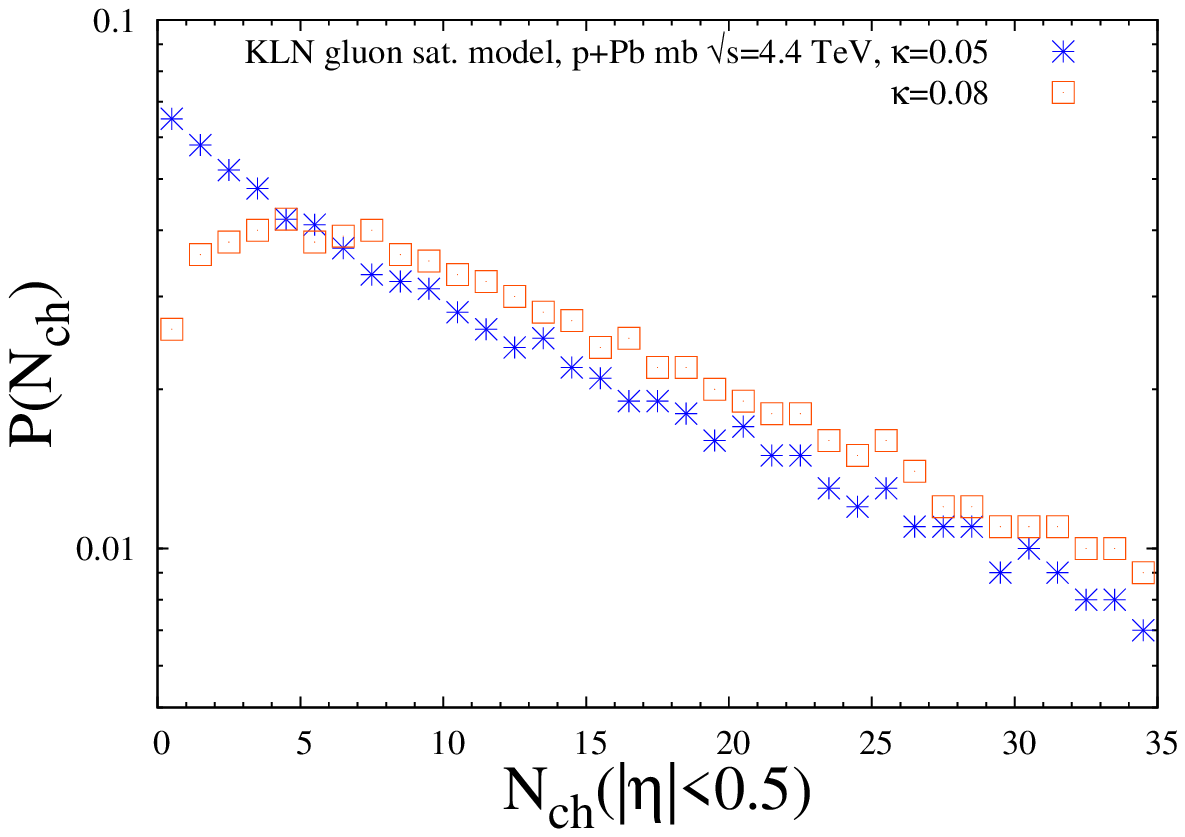}
\end{center}
\caption{
Left: rapidity distribution of charged particles in minimum bias
$p+Pb$ collisions at $W=4400$~GeV. A $\sim 10\%$ overall normalization
uncertainty is not shown explicitly.
Right: Charged particle multiplicity distribution.
} \label{fig:pPb4400mb}
\end{figure}

\begin{figure}[htb]
\begin{center}
\includegraphics[width=0.45\textwidth]{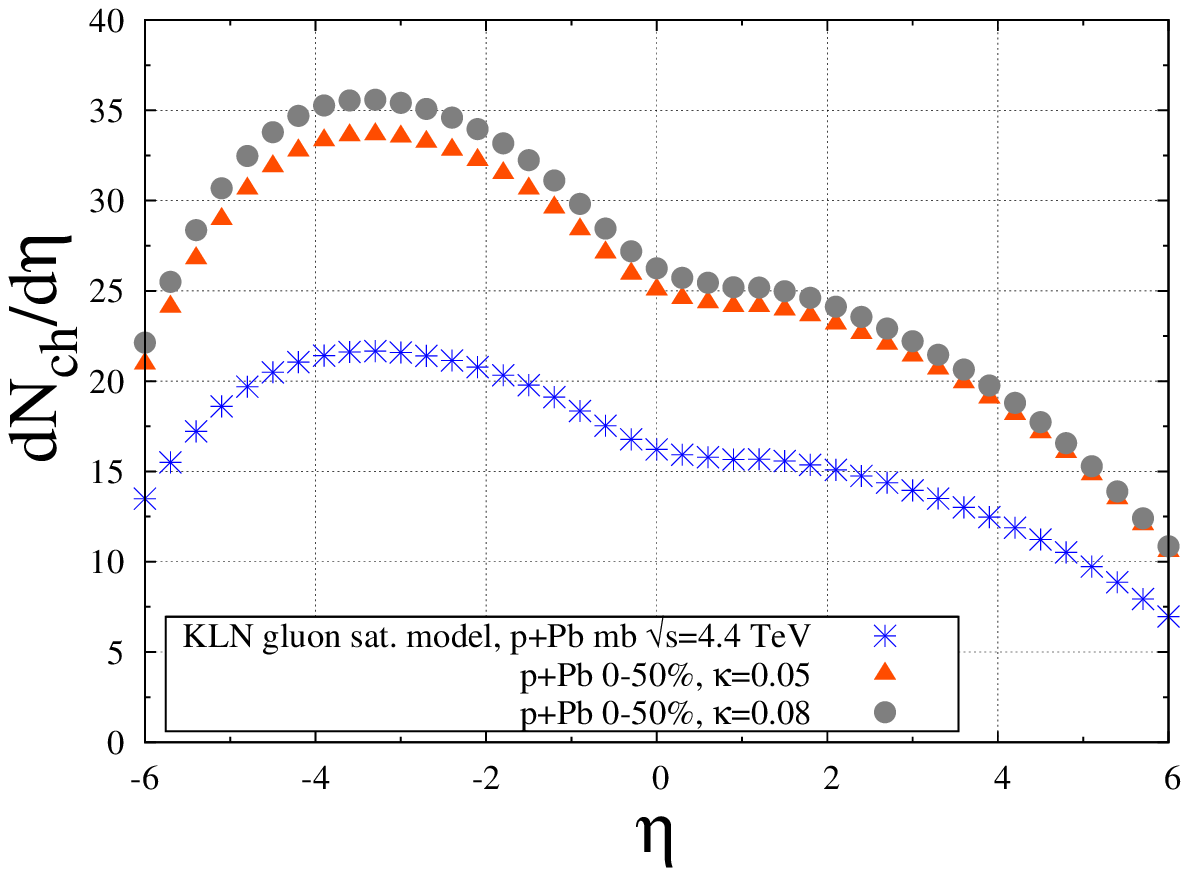}
\end{center}
\caption{
Rapidity distribution of charged particles in
$p+Pb$ collisions at $W=4400$~GeV for minimum bias trigger and for the
0-50\% centrality/multiplicity class. A $\sim 10\%$ overall normalization
uncertainty is not shown explicitly.
} \label{fig:pPb4400centr}
\end{figure}

The centrality dependence of the charged particle multiplicity in Au+Au
collisions at RHIC is shown at Fig. \ref{fig:AA}; the agreement is
very good. The reduction of the intercept of the gluon distribution
(by $\sim 20\%$ in comparison to \cite{Kharzeev:2004if}) allows us to
reproduce well also the LHC Pb+Pb data, see Fig. \ref{fig:AA}.  Figs.\
\ref{fig:pp900},\ref{fig:pp2360},\ref{fig:pp7000} show the comparison
of our model to the pp data from the LHC on charged hadron
multiplicities and multiplicity distributions at $\sqrt{s} = 0.9,\,
2.36$ and 7~TeV, respectively. The agreement is seen to be quite good. Finally,
in Figs.~\ref{fig:pPb4400mb},\ref{fig:pPb4400centr} we present our
predictions for the upcoming p Pb run at $\sqrt{s} = 4.4$ TeV.

\vskip0.3cm To summarize, we have presented updated predictions of the
KLN model for p Pb collisions at the LHC, as well as comparisons to
the RHIC and LHC data on hadron multiplicities and multiplicity
distributions. Clearly, our treatment has been somewhat
model-dependent and involves a few adjustable
parameters. Nevertheless, our model does capture the emergence of a
new dimensionful scale governing QCD interactions at high energies,
and thus expresses in quantitative form the essence of the parton
saturation phenomenon. The comparison of our model to the existing Pb
Pb and the forthcoming p Pb data would also allow to deduce the amount
of additional entropy produced during the evolution of the quark-gluon
fluid in heavy ion collisions~\cite{Dumitru:2007qr}. Our present
treatment assumes no additional entropy production, which corresponds
to the zero viscosity limit; a deviation from our prediction could
signal the presence of viscous effects.  

\vskip0.3cm
A.D.~gratefully acknowledges support by
  the DOE Office of Nuclear Physics through Grant
  No.\ DE-FG02-09ER41620 and by The City University of New York
  through the PSC-CUNY Research Award Program, grant 64132-0042.
  The work of D.K.\ was supported in part by the US Department of
  Energy under Contracts No. DE-AC02-98CH10886 and DE-FG-88ER41723.
 The work of E.L.\ was supported in part by the  Fondecyt (Chile)
 grant 1100648.
{The work of Y.N.\ was partly supported by
Grant-in-Aid for Scientific Research
No.~20540276.}

\newpage
\vskip0.3cm {\bf Note added Oct.\ 19, 2012:}\\ 

\begin{figure}[htb]
\begin{center}
\includegraphics[width=0.45\textwidth]{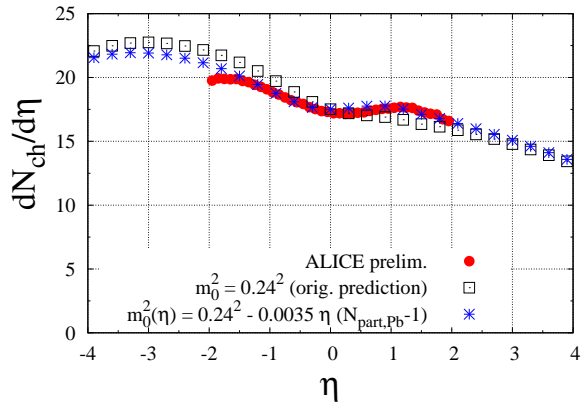}
\end{center}
\caption{Rapidity distribution of charged particles in minimum bias
  $p+Pb$ collisions at $W=5000$~GeV. For these curves we have also
  accounted for the boost of the $\eta=0$ lab frame by adding a
  rapidity shift of $\Delta y = \frac{1}{2} \log\; 82/208 \simeq
  -0.4654$ to the right-hand-side of eq.~(\ref{eq:y_eta}). $m_0$
  denotes the numerator from eq.~(\ref{eq:mu_eta}). } 
\label{fig:TunedJ}
\end{figure}

After our paper was published, data from the ALICE Collaboration on
the charged hadron multiplicity in p+Pb collisions at $\sqrt{s} =
5.02$ TeV appeared~\cite{ALICE1210.3615}. The agreement between the
data and our prediction is quite good over the entire pseudo-rapidity
range of the data, $-2 < \eta < +2$. However, while at $\eta =0$ our
prediction essentially coincides with the data, towards the nuclear
fragmentation region, at $\eta \simeq -2$, the prediction deviates
from the data by about $10\%$, see open boxes in Fig. \ref{fig:TunedJ}.  Here we would like to point out that
this small discrepancy is within the uncertainty resulting from the
Jacobian of the transformation from rapidity to pseudo-rapidity, see
eqs.~(\ref{eq:y_eta},\ref{eq:J_eta},\ref{eq:mu_eta}), and that it can be easily
eliminated.

In our paper, we assumed that the value of the parameter $\mu$
(determined by the typical mass and transverse momentum of the
produced hadrons) in $pA$ collisions is the same as in $pp$
collisions, and is given by eq.~(\ref{eq:mu_eta}). A more accurate
approximation is to assume that the typical transverse mass of the produced hadrons 
is determined by the transverse momentum distribution of
the produced gluons. This leads to the assumption (see e.g.\ eq.~(27) in
\cite{Kharzeev:2001gp}) that the parameter
$\mu^2$ decreases when the saturation momentum grows,
i.e.\ that it varies with
the pseudo-rapidity and the participant density in the Pb nucleus as:
\be
\mu^2(\eta) = \frac{0.24^2 - a\, \eta \, \left[
N_{\rm part(Pb)}-1\right]}{\left(0.13 + 0.32\;W^{0.115}\right)^2}
~,
\ee
where $a$ is a parameter reflecting the rapidity dependencies of
saturation momentum and of the typical mass of produced hadrons. The
sign of the second term describes the growth of the saturation
momentum of the nucleus at small $x$, towards the proton fragmentation
region (positive $\eta$). The (presumably small) magnitude of the
parameter $a$ is determined to large extent by non-perturbative
fragmentation phenomena. In Fig.~\ref{fig:TunedJ} we show the result
of our computation with $a=0.0035$. One can see that a small
adjustment of the Jacobian of rapidity-to-pseudorapidity
transformation that is motivated by the expected features of hadron
production leads to near-perfect agreement with the data.

\end{document}